\documentclass[10pt]{iopart}
\usepackage[english]{babel}
\usepackage{hyperref}
\usepackage{amssymb}
\usepackage{graphicx}
\usepackage{comment}
\usepackage[titletoc,toc,title]{appendix}
\usepackage[colorinlistoftodos]{todonotes}
\usepackage[utf8]{inputenc}
\usepackage{siunitx}
\usepackage{bm}
\usepackage{cite}
\DeclareSIUnit\gauss{G}

\expandafter\let\csname equation*\endcsname\relax
\expandafter\let\csname endequation*\endcsname\relax 
\usepackage{amsmath}
 
\usepackage{physics}
\usepackage{leftidx}
\usepackage{doi}

\definecolor{darkCyan}{rgb}{0, 0.5, 0.5}
\definecolor{lightGrey}{rgb}{0.5, 0.5, 0.5}

\newcommand{\figref}[1]{Fig~\ref{#1}}
\newcommand{\refeqn}[1]{eq~\eqref{#1}}
\newcommand{\needsRef}[1]{\textcolor{darkCyan}{[?]}}
\newcommand{\chk}[1]{\textcolor{darkCyan}{\textbf{*}}}

\newcommand{\Rb}[1]{$^{#1}$Rb}
\newcommand{\Potassium}[1]{$^{#1}$K}
\newcommand{\rf}{rf}
\newcommand{\gF}{\ensuremath{g_F}}
\newcommand{\om}[1]{\ensuremath{\omega_#1}}
\newcommand{\Bohrmag}{\ensuremath{\mu_\mathrm{B}}}

\newcommand{\rfpol}{\ensuremath{\sigma^-}}

\begin{document}

\title{Species-selective confinement of atoms dressed with multiple radiofrequencies}

\author{E.~Bentine, T.~L.~Harte, K.~Luksch, A.~Barker, J.~Mur-Petit, B.~Yuen, and C.~J.~Foot}

\address{Clarendon Laboratory, University of Oxford,
	Oxford OX1 3PU, United Kingdom}
\ead{elliot.bentine@physics.ox.ac.uk}
\begin{abstract}

Methods to manipulate the individual constituents of an ultracold quantum gas mixture are essential tools for a number of applications, for example the direct quantum simulation of impurity physics.
We investigate a scheme in which species-selective control is achieved using magnetic potentials dressed with multiple radiofrequencies,
exploiting the different Landé \gF{}-factors of the constituent atomic species.
We describe a mixture dressed with two frequencies, where atoms are confined in harmonic potentials with a controllable degree of overlap between the two atomic species.
This is then extended to a four radiofrequency scheme in which a double well potential for one species is overlaid with a single well for the other.
The discussion is framed with parameters that are suitable for a $^{85}\mathrm{Rb}$ and $^{87}\mathrm{Rb}$ mixture, but is readily generalised to other combinations.

\end{abstract}

\date{\today}
\maketitle

\ioptwocol

\section{Introduction}

Advances in the experimental techniques used to manipulate ultracold atomic gas mixtures have opened new pathways for the exploration of many-body quantum physics~\cite{Modugno2002, Gunter2006, Cetina2016}, thermodynamics~\cite{Catani2009, McKay2013}, and the formation of ultracold molecules~\cite{Kerman2004,PhysRevLett.97.180404}.
Quantum simulation experiments that use mixtures of atomic species promise new insight into the behavior of impurities coupled to larger quantum systems. Experiments in this field have 
observed non-equilibrium dynamics~\cite{Scelle2013,Cetina2016}, polaronic phenomena~\cite{Kohstall2012,Catani2012} and the disruption and localization of phases by scattered impurities~\cite{Ospelkaus2006}, while control over individual impurities has led to the successful doping of cold gases with single atoms~\cite{Spethmann2012,Hohmann2016}.
Using the impurities as a probe of the larger system presents many
prospects for future experimental work, including the observation of impurity decoherence~\cite{1367-2630-11-10-103055}, Markovianity~\cite{1367-2630-11-10-103055,Galve2010,Haikka2011,Addis2013}, and the non-destructive probing of reservoir excitations~\cite{Hangleiter2015} and correlations~\cite{Streif2016}.

Many of these experiments rely on the use of species-selective potentials to give individual control over the constituent species. 
This is often implemented using an optical dipole trap, at a specific wavelength chosen to interact strongly with one species but not the other~\cite{Leblanc2007}. 
For some mixtures, the wavelength required cannot be reconciled with the constraint of a low heating rate, which requires a large frequency detuning of the dipole trap beam to suppress photon scattering.

An alternative method of species-selective confinement is to use radiofrequency (\rf{}) dressed potentials~\cite{Extavour2006}, which trap atoms in a combination of \rf{} and static magnetic fields~\cite{Zobay2004,Colombe2004}.
The resulting potentials have low heating rates, and are extremely smooth and free from defects when the fields are generated by current-carrying wires located far from the atoms~\cite{Merloti2013}.
Species with Landé g-factors (\gF{}) that differ in sign or magnitude can be manipulated independently.
The magnitude of \gF{} determines the splitting between the Zeeman sub-levels in a static magnetic field, and for an inhomogeneous field this determines the location where the applied \rf{} is resonant.
This has been previously exploited to achieve species-selective control of a
 \Rb{87}-\Potassium{40} mixture on an atom chip~\cite{Extavour2006}, where a single \rf{} resonant with the \Rb{87} Zeeman splitting was applied. This formed a double-well potential with a controllable barrier for the \Rb{87} atoms, while only mildly perturbing the \Potassium{40} potential.
 The sign of \gF{} determines the handedness of the circularly polarised \rf{} field that resonantly couples the atomic Zeeman sub-levels. Control of the \rf{} field's polarisation thus provides a handle to independently manipulate states where the sign of \gF{} differs~\cite{Lesanovsky2006}. This was recently used to realize independent control over the ground state hyperfine levels of \Rb{87} in a \rf{}-dressed matter wave-guide~~\cite{1367-2630-18-7-075014}.

The increased versatility of \rf{}-dressed potentials with multiple radiofrequencies was explored in reference ~\cite{Courteille2006}, which also alludes to species selectivity.
Here, we consider a mixture of atomic species irradiated by an \rf{} field comprised of multiple frequencies, where each species is confined in a potential sculpted by specific frequency components.
Manipulations of each species are made through control of the frequencies, polarisations and amplitudes of the dressing field.
We begin by reviewing the general features of \rf{}-dressed potentials, before discussing the species-selectivity of these traps when extended to multiple dressing frequencies. 
We first consider a mixture irradiated with two radiofrequencies. This confines each species in a harmonic well and permits the controllable overlap or separation of these constituents.
With four radiofrequencies we can implement a double well for one species overlapped with a single well for the other.
We study the application of this approach to a mixture of \Rb{85} and \Rb{87} and conclude by outlining further applications of this dual-species system.

\section{Potentials of \rf{}-dressed atoms}

In a static magnetic field $\mathbf{B}$ atoms have eigenenergies $m_F \gF \Bohrmag \abs{\mathbf{B}}$, corresponding to the Zeeman substates $\ket{m_F}$, and labelled by the projection of the atom's magnetic dipole onto the quantisation axis defined by $\mathbf{B}/\abs{\mathbf{B}}$.
The product of the Bohr magneton \Bohrmag{}, the Landé \gF{} factor and the field magnitude $\abs{\mathbf{B}}$ corresponds to the Zeeman splitting in the low-field linear regime. 
The weak-field seeking states, for which $\gF m_F > 0$, can be trapped at local minima of $\abs{\mathbf{B}}$.
For example, in the magnetic quadrupole field
\begin{equation}
\label{eq:quadrupole}
\mathbf{B}\left(x,y,z\right) = B'(x\mathbf{\hat{e}}_x + y \mathbf{\hat{e}}_y - 2 z \mathbf{\hat{e}}_z)
\end{equation}
with field gradient $B'$, atoms are corralled around the field zero at the origin.

\begin{figure}
\includegraphics[width=0.5\textwidth]{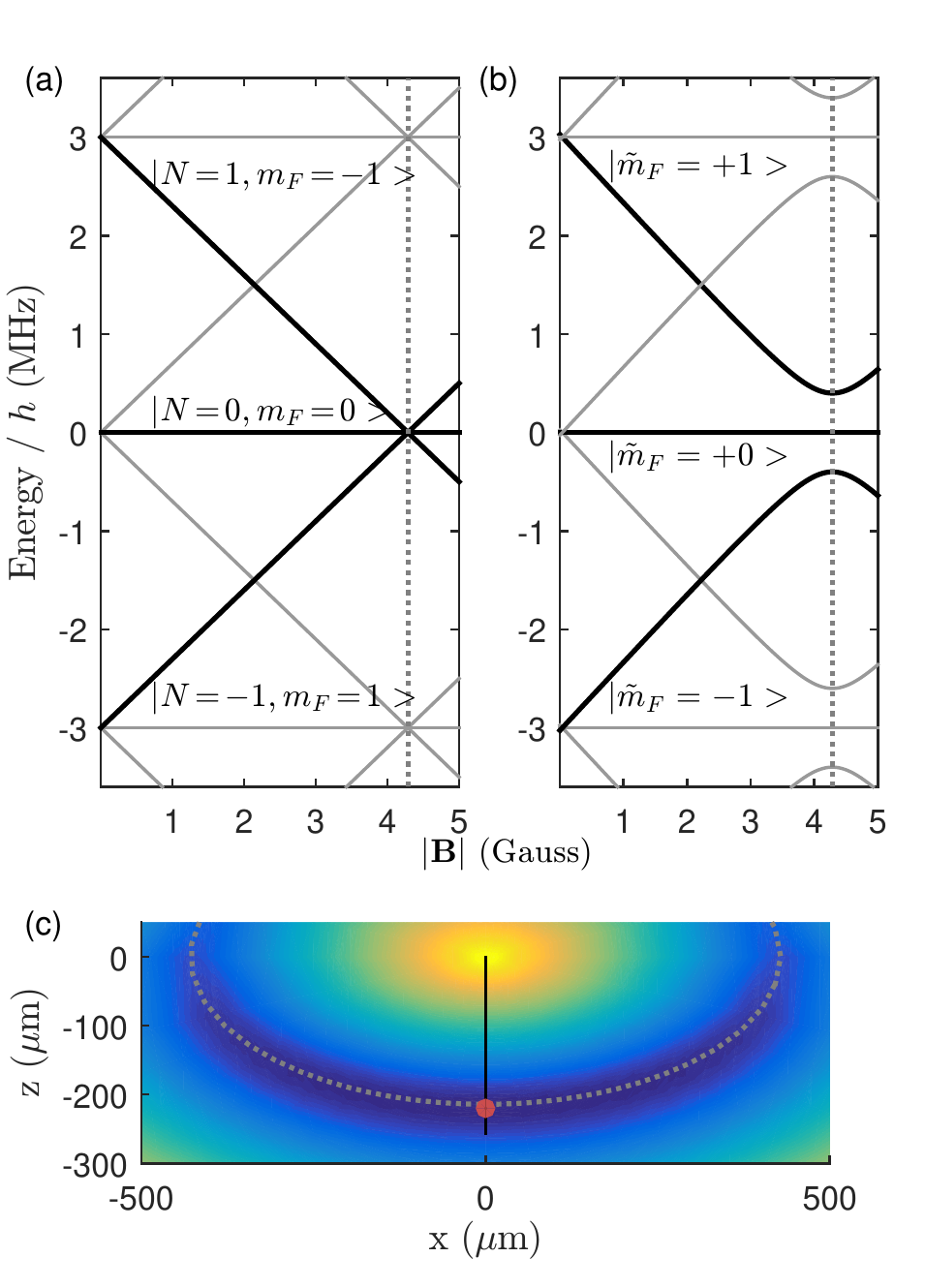}
\caption{
(a) Eigenenergies of the dressed basis states $\ket{N,m_F}$ versus static magnetic field in the absence of interaction for \Rb{87}, $\ket{F=1}$. The \rf{}-photon number $N$ is relative and the zero of energy is arbitrary. One manifold is emphasized in bold.
(b) 
Eigenenergies of the dressed states including the interaction. An avoided crossing forms between states within the same manifold.
(c) Potential energy of atoms in the shell trap as a function of spatial coordinates in the plane $y=0$, including the effect of gravity. A red dot marks the potential minimum, and darker colors indicate a lower energy. The black line depicts the path along which the potential of subfigure (b) is plotted (see (b) for quantitative scale). In all sub-figures $B' = \SI{100}{\gauss\per\cm}$, $\omega = 2 \pi \times \SI{3}{\mega\Hz}$, $\Omega = 2 \pi \times \SI{400}{\kilo\Hz}$,
and the grey dotted line marks the resonance condition.
} 
\label{fig:introToAPs}
\end{figure}

The eigenstates of atoms in a static magnetic field, that are irradiated by a strong field of radiofrequency $\omega/2 \pi$, can be understood using the dressed atom formalism~\cite{Zobay2004,Cohen-Tannoudji}. 
The dressed basis states $\ket{N, m_F}$ are tensor products of the individual Fock states of the \rf{} field mode, $\ket{N}$, and the Zeeman substates of the atom.
In the absence of the atom-photon interaction these states have energy $\gF{} m_F \Bohrmag \abs{\mathbf{B}} + \hbar \omega N$, corresponding to the ladder shown in \figref{fig:introToAPs}a.
In the rotating-wave approximation the interaction of the atom and \rf{} field
couples states within manifolds of constant $\tilde{N} = \mathrm{sign}\left(\gF{}\right)m_F + N$, according to quantum mechanical selection rules which account for the polarisation of the \rf{} field.
Avoided crossings form where the energy of an \rf{} photon is equal to the Zeeman splitting, fulfilling the resonance condition 
\begin{equation}
\label{eq:resonanceCondition}
\gF{} \Bohrmag B = \hbar \omega
\end{equation}
as shown in \figref{fig:introToAPs}b.
The resulting eigenenergies of the magnetic dipole Hamiltonian for the dressed atom take the form
\begin{equation}
\label{eq:dressedAtomEnergies}
U(\mathbf{r}) = \tilde{m}_F \hbar \sqrt{\delta(\mathbf{r})^2 + \Omega(\mathbf{r})^2} + \tilde{N} \hbar \omega
\end{equation}
where $\tilde{m}_F$ labels each dressed eigenstate, $\Omega$ is the Rabi frequency of the dressing field, and $\delta=\omega-\gF{}\Bohrmag{}\abs{\mathbf{B}}/\hbar{}$ is the angular frequency detuning between the applied \rf{} and the Zeeman splitting of the atoms. Atoms are trapped in states for which $U$ has a local minimum at $\delta = 0$.

The eigenstates of the dressed atoms vary spatially, and for an atom to remain trapped as it traverses the avoided crossing it must adiabatically follow the local eigenstate $\ket{\tilde m_F}$.
The precise condition of adiabaticity depends on the atomic motion and trap geometry~\cite{Lesanovsky2006_2}, but for ultracold atoms performing a small amplitude oscillation at frequency $f$ in a harmonic trap, non-adiabatic spin changes are suppressed when $2 \pi f \ll \Omega$.
This process can be the dominant atom loss mechanism in \rf{}-dressed traps and enforces a lower bound on the amplitude of the applied \rf{} field.

For atoms in the static quadrupole field of \refeqn{eq:quadrupole} the resonance condition of \refeqn{eq:resonanceCondition} is satisfied on the surface of an oblate spheroid centered on the origin, forming a `shell' on which atoms are trapped (see \figref{fig:introToAPs}c).
The field direction varies spatially in the static quadrupole, thus changing the relative orientation between the quantisation axis and the uniform \rf{} field.
For an \rf{} field with constant amplitude and polarisation in the laboratory frame,
this introduces a spatial dependence to the Rabi frequency,
often described as a `coupling strength' that varies over the surface of the shell as determined by the selection rules accounting for the polarisation of the \rf{} field.
Nodes are located where the atom-photon interaction falls to zero, corresponding to a vanishing avoided crossing at which adiabatic following cannot be sustained.

Under the influence of gravity atoms collect around the lowest point of this shell, as indicated by the cooler colours of \figref{fig:introToAPs}c. 
An applied \rf{} field that is circularly polarised about the z-axis maximises the atom-photon interaction at this location, driving \rfpol{} transitions between the atomic Zeeman sublevels.
For small oscillations the atoms experience an anisotropic harmonic potential. 
Weak confinement tangential to the spheroid's surface arises from competition between the gravitational potential energy, that pulls atoms to the bottom of the shell, and the spatial variation of coupling strength.
The tight confinement perpendicular to the surface arises from the increase in $\delta$, giving an oscillation frequency $f_z$ for atoms of approximately
\begin{equation}
\label{eq:trapFreq}
f_z = \frac{\gF{} \Bohrmag{} B'}{\pi} \sqrt{\frac{\tilde{m}_F}{\hbar M \Omega{}}}
\end{equation}
for a particle of mass, $M$, as follows from \refeqn{eq:dressedAtomEnergies}.
Higher quadrupole gradients and lower amplitudes of the dressing \rf{} field lead to tighter vertical confinement near the avoided crossing.

\section{Species selectivity with multiple \rf{}s}

Having reviewed the theory of \rf{}-dressed potentials we now explore their species-selectivity when extended to multiple radiofrequencies, which is applicable to any mixture where $\abs{\gF}$ differ.
To illustrate this we consider atoms in a magnetic quadrupole field dressed by 2 radiofrequencies.
 This implements a single potential well for each species with a controllable overlap and separation between them.
 For numerical purposes here we shall consider the mixture of \Rb{85}-\Rb{87} in their lower hyperfine states, labelling the isotopes A and B respectively so that $g_F^A = -1/3$ and $g_F^B = -1/2$.
The discussion is furnished with numerical calculations based on the parameter range of an existing \Rb{87} apparatus that is described in detail elsewhere~\cite{Sherlock2011}. This was recently used to implement \rf{}-dressed potentials with multiple frequency components for a single species~\cite{Harte2017}.
We calculate the eigenstates of atoms dressed by the multicomponent \rf{} field using an exact Floquet formalism~\cite{Shirley1965,Chu1985,Yuen2017}.
 
\begin{figure*}
\includegraphics[width=\textwidth]{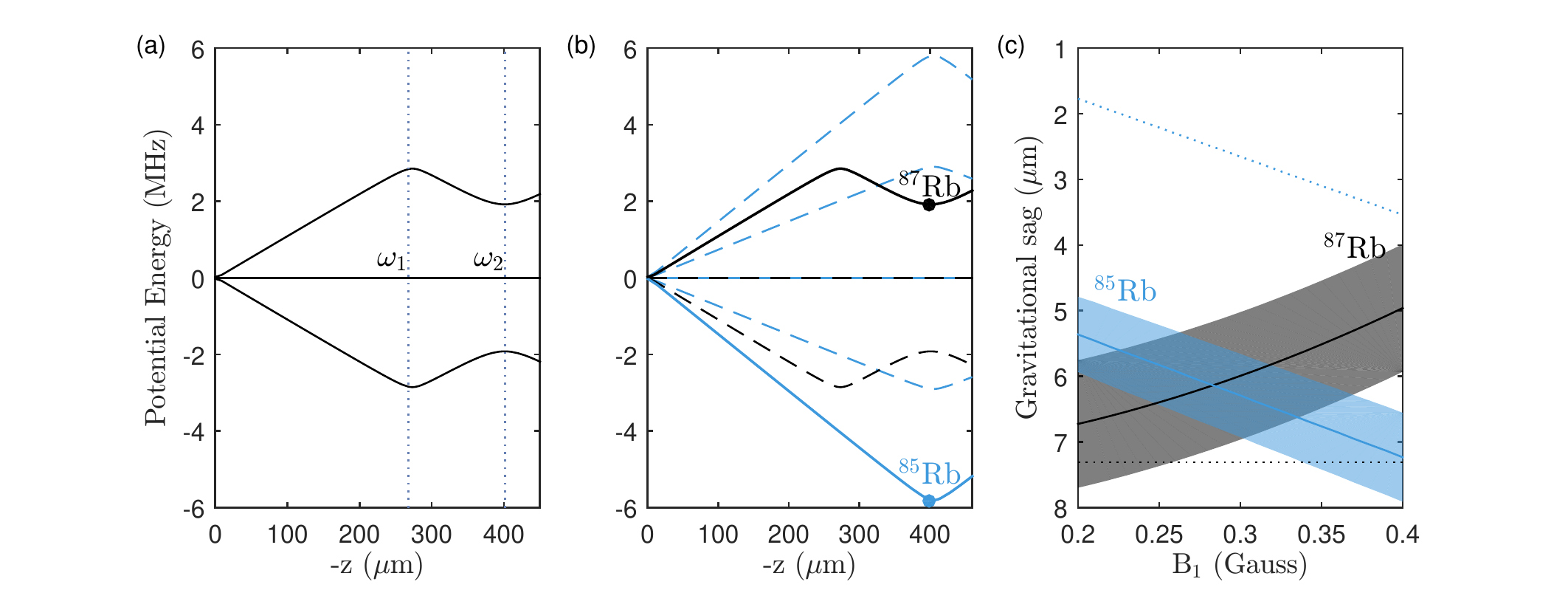}
\caption{
Species-selective \rf{}-dressed potentials with 2 dressing frequencies $\om{1} = \SI{3.0}{\mega\Hz}$, $\om{2} = \SI{4.5}{\mega\Hz}$ confine the isotopes \Rb{85} and \Rb{87} in a quadrupole gradient of \SI{80}{\gauss\per\cm}. For clarity only one portion of the repeating ladder of dressed energy levels is shown.
(a) Eigenenergies of \Rb{87}, $\ket{F=1}$ atoms when dressed by two \rf{}s. Avoided crossings occur close to the light grey dotted lines where the resonance conditions are satisfied.
(b) The eigenenergies of \Rb{87}, $\ket{F=1}$ and \Rb{85} $\ket{F=2}$ are overlaid. Atoms are trapped on the eigenstates plotted as solid lines at the filled circles; other eigenstates are shown as dashed lines.
(c) The spatial separation of the two species in the vertical direction can be dynamically altered by changing the amplitude of a radiofrequency field, for example $B_1$. The vertical axis is the displacement below the unshifted resonance at $z = \hbar \om{1} / 2 g_F^A \Bohrmag B' = \hbar \om{2} / 2 g_F^B \Bohrmag B'$. Solid lines are results from the full multiple-\rf{} calculation while dotted lines show positions from a single-\rf{} calculation to demonstrate the magnitude of the shifts (see text). To provide scale, the filled regions either side of the lines correspond to the harmonic oscillator length of \Rb{85} atoms, and the Thomas-Fermi radius of a Bose-Einstein condensate of $10^4$ \Rb{87} atoms with a radial trap frequency of \SI{9.3}{\Hz}.
}
\label{fig:2rf2species}
\end{figure*}

The \rf{} fields we consider are of the form
\begin{equation}
\mathbf{B}_{\mathrm{\rf}} = \sum_{i=1}^{N} B_i \Big(\cos\left(\omega_i t\right) \mathbf{\hat{e}}_x + \sin\left(\omega_i t\right) \mathbf{\hat{e}}_y \Big)
\end{equation}
with multiple frequency components $i=1,2,...N$ that are circularly polarised in the laboratory frame. We take $N=2$ for our first example of state selective control. Each \rf{} field component $i$ creates an avoided crossing near the location where the species-dependent resonance condition $\hbar \om{i} = \gF{} \Bohrmag{} B(\mathbf r)$ is satisfied, as depicted in \figref{fig:2rf2species}a. 
The distinct values of $\abs{\gF{}}$ for each species cause these resonances to occur at different magnitudes of the static magnetic field, corresponding to resonant spheroids of different radii. 
The frequencies \om{1} and \om{2} are chosen to overlap the resonant spheroids corresponding to these frequencies for \Rb{85} and \Rb{87} respectively.
This is approximately satisfied by frequencies in the ratio $\om{1} / g_F^A = \om{2} / g_F^B$, up to corrections stemming from multi-photon processes that are discussed below.

Atoms of each species can be trapped at this location where the resonant shells overlap by preparing them in the eigenstates that have energy minima here (see \figref{fig:2rf2species}b).
In this configuration, \Rb{85} is near resonance with the field component of frequency \om{1}, and \Rb{87} with that of \om{2}.
Both species can be loaded simultaneously into these wells from a Time-Orbiting Potential trap, as previously demonstrated for \Rb{87} in a \rf{}-dressed potential formed with a single radiofrequency~\cite{Sherlock2011}.
Trapped in this way, species-selective control is attained by modifying the parameters of the corresponding \rf{} field.

For example, each species can be vertically raised or lowered in space by changing the frequency of the associated \rf{} field component, giving a precise control of the relative displacements.
An important experimental consideration, specific to dressing with multiple \rf{}s, is that non-linear processes in the \rf{} generation cause additional frequency components at sum and difference frequencies of the input signal.
These components in turn produce numerous frequencies through higher-order mixing processes. 
Atoms are lost if any of these frequencies become resonant with atomic transitions to untrapped eigenstates during the loading procedure or manipulations of the trap thereafter. This technical limitation constrains the specific choice of dressing \rf{}s, although precise details will depend on the apparatus. 

An alternative approach to raise or lower each species is to keep the radiofrequencies constant, and instead vary the amplitudes of the dressing field components. This method exploits the dependence of the gravitational sag on the vertical trap frequencies for each atomic species.
The radiofrequencies can then be chosen in the exact ratio of the \gF{} factors;
for our example of \Rb{85} and \Rb{87} this reduces the intermodulation products that are present to only integer multiples of $\om{2} - \om{1}$.
In \figref{fig:2rf2species}c the amplitude $B_1$ is varied between $0.2$ and \SI{0.4}{\gauss} while $B_2$ is held at \SI{0.6}{\gauss}, corresponding to trap frequencies $f_z^{85} = 362$ to \SI{259}{\Hz} and $f_z^{87} = \SI{180}{\Hz}$.
The vertical separation changes sufficiently to sweep the clouds between separation and overlap.


Each \rf{} resonance is shifted by the presence of the other \rf{} field component, which displaces the locations of the potential minima.
The magnitude of this effect is comparable to the gravitational sag, as shown in \figref{fig:2rf2species}c.
The leading term is a quadratic shift, which shifts the $i$th resonance by
$\Omega_{j}^2 / 2 \Delta$,
where $\Omega_{j}$ $(j \neq i)$ is the Rabi frequency of the other dressing \rf{} component and $\Delta = \om{2} - \om{1}$ is the frequency separation~\cite{Courteille2006}. These shifts pull the \rf{} resonances closer together, raising the position of the resonance at \om{2} and lowering that of \om{1} by
\begin{equation}
\delta z_i = \frac{\hbar \Omega_{j}^2}{4 \Delta \gF{} \Bohrmag B'}
\end{equation}
This `cross-talk' means that the position of the second species is also affected by the first \rf{} component, and vice versa.
For our example this causes \Rb{87} to rise as \Rb{85} moves downward. If undesired, this effect could be mitigated by changing the amplitude of the other \rf{} component in a complementary fashion.

A relative displacement in the horizontal plane can also be achieved by manipulating the polarisation of the dressing \rf{} field, as was previously demonstrated to tilt a single \rf{} shell trap~\cite{Sherlock2011}. 
In this method a small vertical \rf{} field of the form $\delta \mathrm{B}_i \sin{ \left( \om{i} t + \phi \right)} \hat{\mathbf{e}}_z$ is added to $\mathbf{B}_{\mathrm{\rf{}}}$, breaking the cylindrical symmetry of the coupling strength about $\hat{\mathbf{e}}_z$ for the $i$-th spheroid.
The minimum of potential energy for the atoms is displaced tangentially along the lower surface of the spheroid, in a direction dictated by the phase, $\phi$.
This method has also been used to induce rotation in a cold atom sample~\cite{PhysRevA.85.053401}, where a stirring motion was generated by sweeping the phase to rotate the off-axis potential minimum in a circle about $\hat{\mathbf{e}}_z$.
Employing the same technique here could be used to impart angular momentum to one species while leaving the other unperturbed.

An important aspect for some multiple-species experiments is the ability to adjust the inter-species interaction. 
A widely used approach is to exploit the properties of a Feshbach resonance, where the scattering length between two colliding atoms is modified by a resonance between the entrance channel of the collision and a closed channel corresponding to bound pairs~\cite{Chin2010,PethickAndSmith}. 
The occurrence of these resonances can be controlled using an external magnetic field, a technique that is compatible with optical trapping methods where the field is often a free parameter. 
This method is not well suited to typical \rf{}-dressed traps, which use lower magnetic fields than those required to reach the Feshbach resonances of most species.
We note that although \rf{} fields may be used to control a Feshbach resonance this typically still requires a bias field or high frequencies~\cite{Kaufman2009,Hanna2010,Tscherbul2010}, further complicated by the reduced freedom in these parameters when the \rf{} field provides the confinement mechanism.
Although the inter-species scattering lengths cannot be changed, the interaction between two species can be adjusted by varying their spatial overlap as above.
This also avoids the increased inelastic losses that typically occur near a Feshbach resonance.

For these potentials to be useful for experiments there must also be a way to cool the mixtures to quantum degeneracy.
Evaporative cooling is routinely performed in single \rf{}-dressed potentials where a second weak \rf{} field drives transitions between the dressed states.
For two species held in a two \rf{} trap a third \rf{} field can be applied to selectively evaporate either species.
However, the trajectory of the additional frequency during evaporation must be carefully chosen because of the many possible transitions that arise from multi-photon processes.
Alternatively, atoms can be ejected through a change in their hyperfine state;
the associated change in the sign of \gF{} decouples these atoms from the dressing fields, and they freely exit the trap. 
This is most convenient for constituent species with hyperfine-transition frequencies far removed from each other or the dressing \rf{} frequencies.
Using the precise overlap that is possible with the \rf{}-dressed scheme, the efficiency of sympathetic cooling between two species can be maintained even at very low temperatures. This offers an advantage over purely static magnetic traps, in which the dissimilar gravitational sags of the two shrinking clouds leads to diminished thermal contact when they no longer overlap~\cite{PappThesis}.

\begin{figure*}
\includegraphics[width=\textwidth]{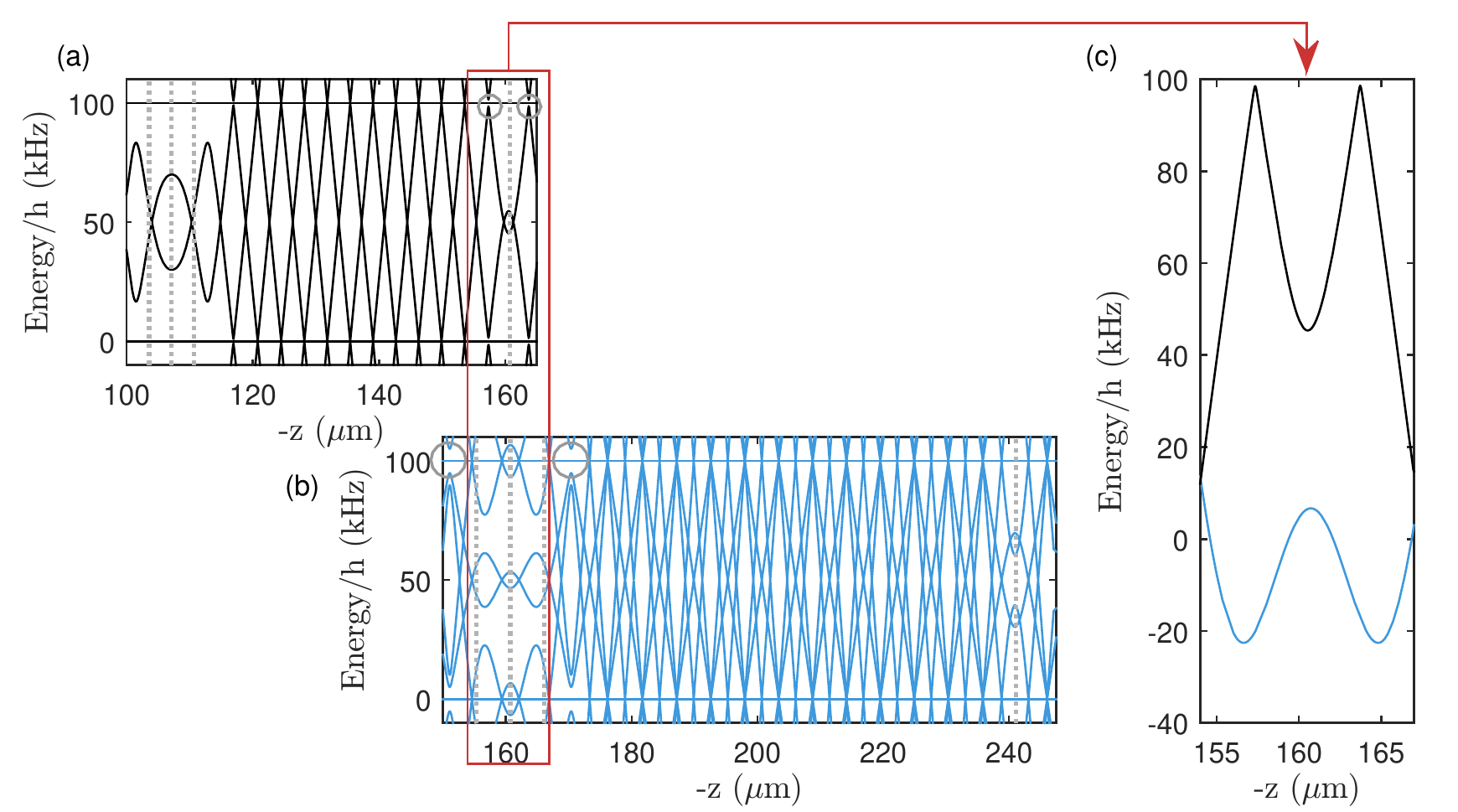}
\caption{
A scheme for creating a double well for \Rb{85} overlapped with a single well for \Rb{87} using an \rf{} field with four frequency components. Both species are in their lower hyperfine levels.
The three frequencies $2.9, 3.0, \SI{3.1}{\mega\Hz}$ create a double well for \Rb{87} centered around \SI{160}{\micro\m} (blue lines). The fourth frequency, \SI{4.5}{\mega\Hz}, creates a single well for \Rb{87} at the same position (black lines). The amplitudes of the \rf{} field components are $B_i = \lbrace 0.1, 0.1, 0.1, 0.065 \rbrace~\si{\gauss}$. Light grey dotted lines depict the unshifted resonances of the dressing frequencies, \refeqn{eq:resonanceCondition}. The dressed eigenstates, including the untrapped states, are shown for (a) \Rb{87} and (b) \Rb{85}, as a function of position along the vertical axis below the quadrupole center. The eigenstates are similar but with a scaling factor of $g_F^A / g_F^B = 2/3$ for position, represented by the shifted and scaled axes. Additionally, there are more levels for \Rb{85} in $F=2$ compared to \Rb{87} in $F=1$. Both confinements have a finite depth caused by small avoided crossings that arise from two-photon processes (grey circles). In (c) we overlay the eigenstates of interest to emphasise the single well/double well structure. The higher trap frequencies of this scheme renders the gravitational sag insignificant here.
}
\label{fig:doubleWellSingleWell}
\end{figure*}

More complex species-selective potentials can be engineered by increasing the number of dressing \rf{} frequencies~\cite{Courteille2006}. 
Shown in \figref{fig:doubleWellSingleWell} are the calculated eigenenergies using an \rf{} field with four frequency components; the lower three frequencies ($2.9$, $3.0$, \SI{3.1}{\mega\Hz}) form a double well potential for \Rb{85} and the highest (\SI{4.5}{\mega\Hz}) a single well to confine \Rb{87}.
These energies are plotted against the position along the vertical axis of the static quadrupole field with gradient $B'=\SI{200}{\gauss\per\cm}$, and so depict the potential energies perpendicular to the surface of the resonant spheroids.

All Rabi frequencies are kept in excess of \SI{30}{\kilo\Hz} to maintain adiabatic following. Similar parameters for an \rf{}-dressed potential with a single frequency have been demonstrated experimentally, with lifetimes of many seconds~\cite{Merloti2013}. 
The tight vertical trap frequencies present in this scheme render the gravitational sag negligible.
Higher order processes generate additional avoided crossings that limit the potential depths for both species.
 
The \Rb{87}, confined in the single well potential, experiences a vertical oscillation frequency of $f_z^{87} = \SI{1.37}{\kilo\Hz}$. While some adjustment to $f_z^{87}$ is possible through the amplitude $B_4$, significantly larger amplitudes cause shifts in the resonances and separate the two species. The height of the barrier is determined by the amplitude of the second \rf{} field, which can be varied dynamically. At a field amplitude $B_2 = \SI{0.1}{\gauss}$ the barrier is $h \times \SI{29}{\kilo\Hz}$ above the double well minima, each well having a harmonic trap frequency of \SI{1.12}{\kilo\Hz}. Increasing the amplitude of $B_2$ to \SI{0.18}{\gauss} lowers the barrier to form a flat potential. Applications of this four frequency species-selective scheme, and the two frequency scheme above, are discussed in the final section.

\section{Mixtures of \Rb{85}-\Rb{87}}

We now consider in more detail the mixture of \Rb{85} and \Rb{87}. 
This combination of species requires a comparatively simple laser scheme;
the required cooling and repumping transitions lie within a span of \SI{6.6}{\giga\Hz}, accessible to a single-frequency laser source via electro-optic modulation of the carrier~\cite{Bonnin2013,Valenzuela:13} or the modulation of current in an injection locked laser diode~\cite{2001RScI...72.2532K}.
Species-selective manipulations of these isotopes using dipole traps are difficult because of the similarity of their optical transition frequencies, but their dissimilar $\abs{\gF{}}$ values makes them well suited to the above method.

The combination of \Rb{85} and \Rb{87} possesses favourable collision properties in a magnetic trap with respect to lifetime and
sympathetic cooling, alongside inter- and intra-species (\Rb{85}) Feshbach resonances~\cite{PhysRevLett.80.2097}.
Experiments have demonstrated tunable interactions in \Rb{85}~\cite{PhysRevLett.85.1795}, the sympathetic cooling of \Rb{85} with \Rb{87}~\cite{Bloch2001}, the formation of ultracold molecules~\cite{PhysRevLett.97.180404} and the controllable miscibility of two condensates~\cite{PhysRevLett.101.040402}.
More recently, this mixture has been used in atom interferometry experiments where a second species can be used to remove common mode noise or test the weak equivalence principle~\cite{Dimopoulos2007,Bonnin2015,Kuhn2014,Zhou2015,STE-QUEST,Williams2016}.

The isotope \Rb{87} is widely employed in Bose-Einstein condensation experiments, whereas \Rb{85} requires more careful consideration.
The majority of experiments that use \Rb{85} rely on a Feshbach resonance to tune the interaction between colliding pairs of \Rb{85} atoms.
At low magnetic fields the negative \Rb{85}-\Rb{85} scattering length of $-460$ Bohr radii~\cite{Blackley2013} leads to collapse; the attractive interaction between particles overwhelms the effective repulsion arising from kinetic energy in condensates of sufficient atom number. This limits the \Rb{85} condensate at low magnetic fields to tens of atoms~\cite{Roberts2001}.
Collapse is not problematic for strongly number imbalanced mixtures, with significantly fewer \Rb{85} atoms than \Rb{87}, as the condensate transition temperatures are also dissimilar, allowing the stable production of a small thermal cloud of \Rb{85} with a \Rb{87} condensate~\cite{PhysRevLett.97.180404}. 

This combination of isotopes is therefore a promising candidate for impurity experiments, where a small number of \Rb{85} impurities interact with a \Rb{87} condensate bath, even in the absence of a tuneable Feshbach resonance. 
At low fields, the \Rb{85} $\ket{F=2,m_F=-2}$ to \Rb{87} $\ket{F=1, m_F=-1}$ scattering length is $213$ Bohr radii~\cite{Kuhn2014,Burke1999}, which permits sympathetic cooling and interaction within the mixture. Further work must be undertaken to determine the inter-species inelastic collision rates of this mixture in the \rf{}-dressed potentials considered here.

\section{Conclusion and Outlook}

We have investigated the utility of multiple-\rf{} dressed potentials to implement species-selective confinement in mixtures of atomic species where $\abs{\gF}$ differ. 
A scheme with two \rf{}s was detailed that allows two species to be spatially separated or brought into contact by manipulating parameters of the dressing \rf{} field.
The complexity of these potentials can be easily extended using additional dressing \rf{} components~\cite{Courteille2006}, and we have shown how four \rf{}s can be used to implement a double well for one species overlapped with a single well for the other.
The isotopes \Rb{85} and \Rb{87} are well suited to this scheme, and promising candidates for population imbalanced impurity experiments.

Many interesting questions can be addressed by mixtures in \rf{}-dressed potentials. 
With each species in an adjustable well, as per the two \rf{} scheme, the system can be used to examine quantum quenches~\cite{Calabrese2006}, the self-trapping of impurities in the condensate~\cite{Cucchietti2006}, or the superfluid drag force exerted on impurities~\cite{Sykes2009}.
The inherent smoothness of dressed potentials generated from macroscopic coils is ideally suited for mechanical rotation experiments, and for species where the mass differs by only a few percent the centrifugal separation of the mixture is minimal. This could be used to realize impurities bound to the vortex cores of a rotating condensate~\cite{Johnson2016}.

The double well for \Rb{85} and single well for \Rb{87} permits an even greater range of investigations. In the absence of tunneling, insights can be made into Markovianity~\cite{1367-2630-11-10-103055,Galve2010,Haikka2011,Addis2013} or the chaotic behaviour of a reservoir coupled to two separate quantum systems (here, the two wells)~\cite{Relano2007}.
Conversely, with tunneling between the wells, demonstrations of non-destructive thermometry and probing of excitations~\cite{Hangleiter2015}, of many-particle correlations~\cite{Streif2016}, and collective decoherence~\cite{1367-2630-11-10-103055} are possible.

This work was supported by the EU H2020 Collaborative project QuProCS
(Grant Agreement 641277) and Spain's MINECO Project No. FIS2015-70856-P.

\section*{References}

\appendix

\bibliographystyle{iopart-num}
\bibliography{Rb8587}

\end{document}